\documentclass[sn-mathphys]{sn-jnl}
\usepackage{hyperref}
\usepackage{amsmath,amssymb,amsfonts,amsthm}

\usepackage{color,bbold}

\newcommand{\N}{{\mathbb N}}

\newcommand{\cH}{{\mathcal H}}
\newcommand{\cA}{{\mathcal A}}

\newcommand{\be}{\begin{equation}}
\newcommand{\ee}{\end{equation}}
\newcommand{\beq}{\begin{eqnarray}}
\newcommand{\eeq}{\end{eqnarray}}

\def\dagg{^\dagger}
\jyear{2023}%

\begin{document}

\title[On the tensorial structure of general covariant quantum systems]{\centering{On the tensorial structure of \\ general covariant quantum systems}}


\author[1]{\fnm{Gabriel M.} \sur{Carral}}

\author[2]{\fnm{I\~naki} \sur{Garay}}

\author[3]{\fnm{Francesca} \sur{Vidotto}}

\affil[1]{\orgdiv{Department of Applied Physics}, \orgname{University of Santiago de Compostela}, \orgaddress{\city{ \postcode{15782} Santiago\,de\,Compostela}, \state{Galicia}, \country{Spain}}}

\affil[2]{\orgdiv{Department of Physics and EHU Quantum Center}, \orgname{University\,of\,the\,Basque\,Country  UPV/EHU}, \orgaddress{Barrio Sarriena s/n, \postcode{48940} \city{Leioa}, \state{Basque Country}, \country{Spain}}}

\affil[3]{\orgdiv{Department of Physics and Astronomy, Department of Philosophy, and Rotman Institute}, \orgname{Western University}, \orgaddress{ \postcode{N6A\,5B7}, \city{London}, \state{Ontario}, \country{Canada}}}



\abstract{
The  definition of a quantum system requires a Hilbert space, a way to define the dynamics, and an algebra of observables. The structure of the observable algebra is related to a tensor product decomposition of the Hilbert space and represents the composition of the system by subsystems. It has been remarked that the Hamiltonian may determine this tensor product structure. Here we observe that this fact may lead to questionable consequences in some cases, and does extend to the more general background-independent case, where the Hamiltonian is replaced by a Hamiltonian constraint.  These observations reinforce the idea that specifying the observables and the way they interplay with the dynamics is essential to define a quantum theory. We also reflect on the general role that system decomposition has in the quantum theory. 
}

\maketitle

\section{Introduction}
What are the minimal mathematical ingredients that define a quantum theory?  Different textbooks take different perspectives in presenting quantum physics. Some take a (non-commutative) observable algebra $\cal A$ as the basic ingredient \cite{Dirac}, while others focus on the Hilbert space $\cal H$ and the dynamics defined, say, by a Hamiltonian $H$ \cite{Carroll:2021aiq}.  Here we focus on a specific question: is a quantum theory fully defined by its Hilbert space and its dynamics, or does the observable algebra need to be independently specified?   The first possibility has been explored in \cite{Cotler:2017lfs,Carroll:2018mde}, on the basis of some interesting theorems that suggest that an observable algebra may be implicitly defined by a tensorial decomposition of the Hilbert space selected by the Hamiltonian.  Here we further discuss this possibility, pointing out some potential difficulties. 

A tensorial decomposition $\cal T$, or \emph{Tensor Product Structure} (TPS), is the decomposition of a Hilbert space into the tensor product of $N$ factors.  Physically, this describes the partition of a system into subsystems.  An observable associated to a single subsystem acts only on the corresponding factor.  There is therefore a strict connection between the TPS and the structure of the observable algebra: the TPS reflects the structure of the sub-algebras of the observables of the individual components \cite{Zanardi:2004tps}. Is the decomposition into subsystems determined by the Hilbert space and its dynamics, or must it be independently specified, to describe a quantum system?  

A preferred TPS $\cal T$ and a related observable algebra can be extracted from the couple $({\cal H}, H)$ by requiring that $H$ is \emph{local} with respect to $\cal T$, in a sense that we detail below \cite{Cotler:2017lfs}. The relevance of these results in the general quantum gravity context has been argued in \cite{Bao:2017fhs}.

Here we point out an example in which trying to read the TPS from the Hamiltonian has the result of hiding the relevant physics of a system, suggesting that the role of the observable algebra remains crucial. 

This appears even more relevant in quantum gravity. The form of a quantum gravity theory is very different from a $({\cal H}, H)$ structure, because of the absence of a canonical time and a canonical Hamiltonian. Instead, the dynamics is defined relationally by a constraint \, $C\!=\!0$ \, defined over an \emph{extended} Hilbert space $\cal K$ \cite{Rovelli:2004book}.  In  this more general  setting the selection of a TPS by the dynamics  fails, pointing again to the need of specifying the system's partition or the observable algebra, in order to have a meaningful quantum theory.   

Partitions play a ubiquitous role in physics.  The very definition of observability appears to be tied to the split of the global system into an observed/observing systems, or system/apparatus.  Partitions are at the root of the emergence of specific values for physical variables, according to interpretations that range from Copenhagen (where the focus is on the system/apparatus split), to Many Worlds (where observables take only value in a branch and relative to another system), Relational Quantum Mechanics (where observables are only relative to partitions), and Q-Bism (which requires to distinguish the holder of information from the object of knowledge). Even the distinction between the clock variable and the dependent dynamical variables on which non relativistic (and special relativistic) mechanics is based, relies on such a split.   In the second part of the paper, we discuss this ubiquitous role of partitions in physics.

\section{Tensor product structure}\label{TPSSection}

If $\cH_1$ is the Hilbert space of the states $\psi_1$ of a system $S_1$, and $\cH_2$ is the Hilbert space of the states $\psi_2$ of a system $S_2$, then generic states of the composite system $S_1\cup S_2$ are not described by couples $(\psi_1,\psi_2)$ as one could naively expect: they are described by states in the Hilbert space $\cH=\cH_1\otimes \cH_2$. This fact, which is one of the pillars of quantum theory (perhaps not always sufficiently emphasized in introductory classes), is the root of entanglement, quantum correlations and von Neumann entropy. 

Consider the opposite question: given a quantum theory defined by a Hilbert space $\cH$, how do we know if it describes a composite system, and, if so, what is the corresponding decomposition  
\be
\cH=\otimes_i\cH_i\ \ ?
\label{factorization}
\ee  An answer can be found in the structure of the observables' algebra.  As clearly discussed in  \cite{Zanardi:2004tps}, the notion of decomposition of a quantum system into components is encoded into the set of the observables that describe it, which is to say in the way we access it.  The observables that define the $i$-th component of the system act only on $\cH_i$. They form an algebra $\cA_i$. Acting on $\cH$, the different algebras $\cA_i$ form a set of subalgebras that commute with one another, and have only the identity in their intersection: 
\be
\mathcal{A}=\otimes^n_{i=1} \mathcal{A}_i, \ \ \  [\mathcal{A}_i, \mathcal{A}_j]=0\ {\rm for}\ i\neq j,\ \ \  \mathcal{A}_i \cap \mathcal{A}_j=\mathbb{1}.
\label{algebras}
\ee
This decomposition of the observable algebra fixes the tensorial structure \eqref{factorization} of the Hilbert space, and viceversa. The partitioning of a given Hilbert space is thus determined by the actually experimentally accessible observables. 

A different perspective on the root of the decomposition \eqref{factorization} has been suggested in \cite{Cotler:2017lfs} and explored in \cite{Carroll:2018mde} and \cite{Carroll:2020gme}. The idea is that the decomposition is determined by the dynamics, and the structure of the algebra follows. If so, the core of the quantum theory can be simply taken to be the couple $(\cH, H)$, where $H$ is the Hamiltonian, and the structure described above is determined by this couple. 

This possibility is opened by an intriguing observation made in \cite{Cotler:2017lfs}.  The typical physical dynamics that we know are highly local with respect to a Hilbert space factorization of the form \eqref{factorization} in the following sense. The Hamiltonian contains terms diagonal in the factorization and terms that couple {\em a few} factors at most. For instance, the Coulomb interaction terms couple only two particles among themselves, and the Hamiltonian of a scalar field on the lattice couples only sites that are a few steps away from one another.   Formally,  a generic operator can be written as
\be
H=\sum_i H_i +  \sum_{i>j} H_{ij} + \sum_{i>j>l} H_{ijl} + ...
\ee  
where $H_i$ acts only on $\cH_i$,  $H_{ij}$ acts only on $\cH_i\otimes \cH_j$, and so on.   We say that an operator is $k-$local with respect to the factorization \eqref{factorization} if the above expansion has only the first $k$ sums.   That is, $H$ is 1-local if it is a sum of terms acting on single factors, it is 2-local if in addition has terms coupling \emph{two} factors, and so on.  

Now, the first key observation is that, for low $k$, a generic operator is not $k$ local with respect to any decomposition of the Hilbert space.  This can be intuitively seen as follows, for finite (but many) dimensional Hilbert spaces.  Take $k=2$, and consider a decomposition in $m$ factors each of dimension $n$. The total space has then dimension $d=n^m$. Since the Hamiltonian is defined by its (real) spectrum, the space of Hamiltonians has dimension $n^m$. On the other hand, the dimension of the space of all hermitian operators of the form $H=\sum_i H_i +  \sum_{i>j} H_{ij}$ is (disregarding permutations) $mn+(n(n-1)/2)n^2$, which for large $n$ and $m$ is clearly smaller that $n^m$ (see \cite{Cotler:2017lfs} for details).  For low $k$, and a sufficiently large number of dimensions and components,  a generic operator is not $k$-local with respect to any factorization.   

Next suppose that a Hamiltonian $H$ is given and happens to be $k$-local with respect to a given factorization.  Could the same Hamiltonian $H$ also be $k$-local with respect to a different factorization?  The question is investigated rigorously in \cite{Cotler:2017lfs} and the answer appears to be negative under quite generic conditions, on which we do not enter here. These results have been proven only for systems with a finite dimensional Hilbert space, but it seems sensible that they remain valid in the case of infinite dimensional systems \cite{Cotler:2017lfs}.

All this leads to an intriguing hypothesis: that a (suitable) couple $(\cH, H)$ may be sufficient to determine a preferred factorization: the one that minimizes the $k$ of the $k$-locality of $H$.  In physical terms, the hypothesis is that what we call ``components" of a systems are those into which the system can be decomposed in such a way that the dynamics couple only a few of them at the time.  

Since, as we have seen, the decomposition selects a preferred family of observable subalgebras, one is tempted to make the hypothesis that a quantum theory could be entirely defined by the couple  $(\cH, H)$, without need to independently specify the observable algebra \cite{Carroll:2018mde}. 

Is this physically reasonable?  We now start discussing this hypothesis by looking at a simple quantum system that might shed some doubt on it.

\subsection{Two TPS's for the double quantum oscillator}

Consider a simple example where we see effect of having two different factorizations.  The example also helps us introduce some ingredients that we shall use later on.  Consider two coupled harmonic oscillators. The system is governed by the Hamiltonian 
\begin{equation}\label{Hamop0}
H=\frac{p_1^2}{2m_1}+\frac{p_2^2}{2m_2}+\frac{m_1 \omega_1^2 x_1^2}{2} +\frac{m_2 \omega^2_2 x_2^2}{2}+\frac{\gamma}{2}(x_1-x_2)^2. 
\end{equation}
In presenting the Hamiltonian in this form, we are giving a preferred set of observables: the positions $x_1$ and $x_2$ of the oscillators and their momenta $p_1$ and $p_2$.  The two operators $x_1$ and $p_1$ define the subalgebra $\cA_1$ and  the two operators $x_2$ and $p_2$ define the subalgebra $\cA_2$. These satisfy \eqref{algebras}. The corresponding factorization of the Hilbert space $\cH=\cH_1\otimes \cH_2$ gives the factorization of the coupled system into the two oscillators. This Hamiltonian is clearly 2-local with respect to this TPS. 

As well known, it is possible to diagonalize the Hamiltonian \eqref{Hamop0} by performing a linear canonical transformation of the positions and their momenta \cite{quant-ph/0403184}. 
Using $m=\sqrt{m_1m_2}$, we define the normal mode variables 
\begin{eqnarray}
X_1&=&\lambda x_1\cos\alpha+\frac{1}{\lambda}x_2\sin\alpha,\\
X_2&=&-\lambda x_1\sin\alpha+\frac{1}{\lambda}x_2\cos\alpha,\\
P_1&=&\frac{1}{\lambda}p_1\cos\alpha+\lambda p_2\sin\alpha,\\
P_2&=&-\frac{1}{\lambda}p_1\sin\alpha+\lambda p_2\cos\alpha,
\end{eqnarray}
where $\lambda =(m_1/m_2)^{1/4}$ and 
$\tan 2\alpha =\frac{2\gamma/m}{\omega^2_2-\omega^2_1+\frac{\gamma}{m}\frac{m_1-m_2}{m}}$.
In terms of these, the Hamiltonian reads 
\begin{equation}\label{Hamdiagop}
H=\frac{1}{2m}(P^2_1+P^2_2)+\frac{1}{2}m(\Omega^2_1 X^2_1+\Omega^2_2 X_2^2),
\end{equation}
where the new frequencies $\Omega_j$ are 
{\allowdisplaybreaks
\begin{eqnarray}
\hspace*{-5mm}\Omega_1^2 &=&\omega_1^2 \cos^2\! \alpha+\omega_2^2 \sin^2\! \alpha\!+\!\frac{\gamma}{m}
\left(
\lambda\sin\alpha-\!\frac1\lambda \cos\alpha
\right)^{\! 2}\!\!,\\
\hspace*{-5mm}\Omega_2^2 &=& \omega_1^2 \sin^2\! \alpha+\omega_2^2 \cos^2\! \alpha\!+\!\frac{\gamma}{m}
\left(
\lambda\cos\alpha+\!\frac1\lambda \sin\alpha
\right)^{\! 2}\!\!.
\end{eqnarray}
}
Now, the Hamiltonian \eqref{Hamdiagop} describes two uncoupled harmonic oscillators: the normal modes of the system. What we have done is to introduce a new TPS 
\begin{equation}
 \mathcal{H} = \mathcal{H}'_1 \otimes \mathcal{H}'_2,
\end{equation}
with respect to which the same Hamiltonian is $1$-local.   The Hilbert subspaces $\mathcal{H}'_1$ and $\cH_2'$ correspond to the two normal modes. The corresponding subalgebras are given by the position and momentum of these normal modes, that is $\{(X_1, P_1), (X_2, P_2)\}$.

This simple example raises some worries for the hope of reading everything in the Hamiltonian alone.  The physics of two coupled oscillators includes interesting phenomena like beats (interferences between the frequencies), slow oscillation of the energy from one oscillator to the other if the coupling is small, degeneracy split, and so on.   All this rich phenomenology disappears entirely if we only look at the normal modes variables. More specifically, all these phenomena are described by, and pertain to, the original variables $(x_1,x_2)$ and are lost in terms of the normal modes variables $(X_1,X_2)$. We can of course recover these phenomena in  terms of normal modes variables, but only at the price of knowing the relation between the two sets of variables, which is exactly the information that the algebra of observables has in addition to the information contained in the Hamiltonian. We loose all this if we assume that all relevant physical information is coded into the sole Hamiltonian. 

As we shall see, this is the core of the problem that we shall find in the general case in the next section. 

Physically, the Hamiltonian describes the physical interactions between the system's components. The observables in the observable algebra describe possible interactions between the system and the external observers.

\section{General covariant systems}\label{GCFsection}

All known fundamental systems can be described in the formalism of general covariant mechanics.   This is a simple extension of conventional Hamiltonian mechanics; it includes conventional Hamiltonian mechanics as a special case.  The extension is necessary to describe relativistic gravitational systems, for which the conventional formalism is unsuitable.  

The main difference between  conventional and general covariant formalisms is that the first is based on the specification of a preferred `time' variable $t$; while the second treats all variables (including any independent or `time' variable) on the same ground.  The first describes dynamics as evolution of physical variables in $t$.  The second describes dynamics as evolution of physical variables with respect to each other, without specifying an `independent' one \cite{Rovelli:2004book}. 

In the classical case, the general covariant mechanics of a system with $N<\infty$ degrees of freedom can be defined as follows  \cite{Rovelli:2004book,Rovelli:2014book}.   The variables $q_a$ with $a=1,...,N+1$ that describe the system define an extended configuration space $\cal C$. Its tangent space is the extended phase space $\Gamma$. Dynamics is defined by a function $C$ on $\Gamma$ as follows.   On the subspace $C=0$, the orbits generated by the Hamiltonian flow of $C$ are the ``motions" of the system.  Each motion establishes relations between the variables $f(q_a)=0$, which yields the physical predictions. 

The special case formed by conventional Hamiltonian mechanics is obtained when one of the $N$ variables is singled out and called $t$, so that we write the variables as $q_a=(q_n,t)$ with $n=1,...,N$, and $C$ has the form 
\be \label{split}
C=H+p_t, 
\ee
where $H$ is a function of the $q_n$ and their momenta and $p_t$ is the momentum conjugate to $t$.  In this case, the orbits are monotonic in $t$ and every motion $f(q_n,t)=0$ defines the set of functions $q_n(t)$. This is the conventional definition of dynamics in terms of evolution with respect to a preferred canonical time variable. 

The quantities $q_a$ (that is, both the variables $q_n$ and $t$) are called partial observables  \cite{Rovelli:2001po}.   

In the Lagrangian formalism, a general covariant system is described by a Lagrangian ${\cal L}(q_a,\dot q_a)$ that is general covariant in its (unphysical, or gauge) evolution parameter $\tau$.   The Legendre transform of the system gives a vanishing canonical Hamiltonian and $C$ as a first class constraint.  The solutions of the Euler-Lagrange equations are the motions, expressed using a parametrization $q_a(\tau)$. The equations of motion are invariant under a (`gauge')  arbitrary smooth invertible reparametrization $\tau\to\tau'(\tau)$, which indicates that the quantity $\tau$ is an irrelevant gauge variable and the physics is only given by the graph of these motions, namely as a set of relations between the $q_a$, as above.  

The formalism can be extended to field theory. General relativity, with any matter coupling, can be treated in this general form, without a special variable having to be picked as the canonical time variable.   This is how observational and experimental relativistic gravity work. 

In the quantum domain, the general covariant formalism can work as follows  \cite{Rovelli:2004book,Rovelli:2014book}.  The partial observables $q_a$ and their conjugate momenta form a non commutative algebra. This can be represented on a Hilbert space $\cal K$, called the  `extended' quantum state space.    The constraint $C$ defines a constraint operator, which we indicate with the same notation $C$.  Here we consider for simplicity only systems where $C$ is a single function, and zero is in the spectrum of $C$, but the following can be generalized to the case of many constraints and zero in the continuous spectrum. The solution of (the `Wheeler de Witt equation') 
\be \label{WdW}
C\psi=0, 
\ee
with $\psi\in{\cal K}$, form the null eigen-space $\cal H$ of $C$. The orthogonal projector $P:{\cal K}\to {\cal H}$ can be written formally as 
\be
P=\int_{-\infty}^{\infty} d\tau \ e^{iC\tau}.
\ee
Given the eigenvalues $q_a$ of a complete set of commuting partial observables, the dynamics is defined by the transition amplitudes 
\be
W(q_a,q'_a)=\langle q_a \vert P\vert q_a'\rangle.
\ee
See  \cite{Rovelli:2004book,Rovelli:2014book} for details.  In the conventional case where $C=H+p_t$, it is easy to see that \eqref{WdW} becomes precisely the Schr\"odinger equation, and a straightforward calculation  \cite{Rovelli:2004book}  gives 
\be
W(q_n,t,q'_n,t')=\langle q_a \vert e^{iH(t-t')}\vert  q_a'\rangle,
\ee
which can be recognized as the usual transtion amplitudes that define the quantum dynamics. 

Notice that the conventional case imposes a TPS: it splits $\cal K$ into the tensor product of the Hilbert space $\cal H$ and a `Hilbert space of the clock', where the operators $(t,p_t)$ are defined. In this sense, the choice of a time variable is given by a TPS, or a partition of the full system into `clock' and `rest' \cite{Carroll:2018mde, Colosi:2003si,Milburn2005}. 
In this sense, time is a type of TPS.

The so called ``Clock Ambiguity'' pointed out in \cite{Albrecht2012, Albrecht:2007mm} consists in the fact that a partition like \eqref{split} can be done in different manners.  Contrary to what is sometimes stated, this is not a difficulty; it is simply a reflex of the fact that in general relativistic physics there is no canonical time variable, and we can choose different variables as the independent one.  If we choose an independent variable that does not give rise to the split \eqref{split}, we obtain a non unitary generalization of the Schr\"odinger equation \cite{timeinQG}.  This lack of unitarity does not jeopardize the probabilistic interpretation, which remains valid: it is only the time evolution that is altered, because the variable chosen as `clock' does not run monotonically in $[-\infty,\infty]$.
 
\subsection{Two oscillators without time}

A simple example of a general covariant system that does not admit a formulation as a deparametrized system was studied in detail in  \cite{Colosi:2003si,Rovelli:2014book}. This is obtained by taking the operator  
\eqref{Hamdiagop} as the constraint defining the quantum dynamics of a system with a single degree of freedom, as follows. We consider the system defined by the Wheeler De Witt equation
\begin{equation}\label{Ham0diagop}
C\psi=\left(\frac{1}{2m}(P^2_1+P^2_2)+\frac{1}{2}m(\Omega^2_1 X^2_1+\Omega^2_2 X_2^2)-E\right)\psi=0,
\end{equation}
where $E$ is a constant that we shall fix in a moment. 
Making use of the creation and annihilation operators $a_j$ and $a\dagg_j$ defined in the usual way
\begin{equation}
a_j\!=\!\sqrt{\frac{m\Omega_j}{2\hbar}}\left(\!X_j\!+\frac{iP_j}{m\Omega_j}\!\right),\quad
a^{\dagger}_j\!=\!\sqrt{\frac{m\Omega_j}{2\hbar}}\!\left(\!X_j\!-\frac{iP_j}{m\Omega_j}\!\right),
\end{equation}
we can write the constraint \eqref{Ham0diagop} in terms of the number operator $N_j\equiv a^{\dagger}_ja_j$ in the following way:
\begin{equation}\label{eq5}
C=\left(N_1+\frac{1}{2}\right)\Omega_1+\left(N_2+\frac{1}{2}\right)\Omega_2 -E. 
\end{equation}
 In the basis of the eigenvectors of the number operator,  the Wheeler-de Witt equation \eqref{WdW}  takes the form 
\begin{equation}\label{eq8}
C\lvert n_1 n_2\rangle=\biggl[\biggl(n_1+\frac{1}{2}\biggl)\Omega_1+\biggl(n_2+\frac{1}{2}\biggl)\Omega_2-E\biggl]\lvert n_1 n_2\rangle=0.
\end{equation}
This restricts on the allowed values that $E$ can take for the system to have solutions. Once this is chosen, the maximum value of $n_1+n_2$ is constrained, and $\cal H$ is a finite dimensional Hilbert space.
Consider for simplicity the case $\Omega_1=\Omega_2=1$, the Wheeler-de Witt equation \eqref{eq8} reduces then to 
\begin{equation}\label{eq9}
n_1+n_2=E-1\equiv N.
\end{equation}
Taking into account equation \eqref{eq9} we relabel the basis states as
\begin{equation}
\vert n\rangle \equiv \vert n,N-n\rangle.
\end{equation}
These states form a basis of the subspace $\mathcal{H}$ of the original Hilbert space $\cal K$ that solve the Wheeler de Witt equation.   It has dimension 
$N+1=E\in\N$. The projector from $\cal K$ to $\cal H$ is 
\begin{equation}
P=\sum_{n=0}^N \vert n\rangle\langle n\vert .
\end{equation}
The  Hamiltonian constraint in this basis is of course vanishing, by construction. 
The `time evolution' with respect to $\tau$ generated by this constraint vanishes for any observable 
\begin{equation}\label{quantevoleq}
\frac{dO}{d\tau}=\frac{i}{\hbar} [O,H]=0,
\end{equation}
as it should, since now this is an unphysical gauge.   Here $O$ is some observable and $\tau$ is the non-physical parameter of the general covariant formalism described in section \ref{GCFsection}. 

Equation \eqref{quantevoleq} does not mean that there is no change or evolution on the two oscillator system. On the contrary, it implies that we are working in a formalism without a preferred time. Evolution will be expressed in terms of correlations between partial observables.   In fact, standard transition amplitudes can be defined
\be
W(x'_1, x'_2; x_1, x_2)=\langle x'_1, x'_2\vert P\vert x_1, x_2\rangle.
\ee
These can be written explicitly in terms of the standard energy eigenstates of the harmonic oscillator in the position basis $H_n(x)$
\be
W(x'_1, x'_2; x_1, x_2)=\sum_{n=0}^N \overline{H_{n}( x'_1) H_{N\!-\!n}(x'_2)} H_{n}(x_1) H_{N\!-\!n}(x_2).
\ee
The transition amplitudes $W(x_1, x_2; x'_1, x'_2)$ are analogous to the usual transition amplitudes of a single degree of freedom system $W(x, t; x', t')$, with the only difference that there is no a priori distinction between an ``independent" variable $t$ and a ``dependent" variable $x$: the two are treated on equal ground.  For the definition of the probability in terms of these generalized amplitudes, see   \cite{Rovelli:2004book}.

\subsection{TPS in general covariant systems}\label{CTPSSection}

Can the idea of defining a physical TPS from the dynamics be extended to general covariant systems?  In the case in which the dynamics is defined by a Hamiltonian, it is the spectrum of the Hamiltonian that carries the relevant information, as this is the only information the Hamiltonian carries, in the absence of other structures.  Clearly, this cannot be extended to a general covariant system.   Here the dynamics is defined by a Hamiltonian constraint operator.  The spectrum of this operator restricted in the physical Hilbert space $\cH$ is zero, and the rest of the spectrum is physically irrelevant, because the only role of the operator is to select out $\cal H$.  The information about the tensorial structure must be given by something else. It cannot be just in the Hamiltonian constraint.  

In fact, this information will be given by the other observables apart of the Hamiltonian. The two examples with the double pendulum given above clearly point to where the relevant information is: in the TPS of the extended Hilbert space $\cal K$ which is defined by the algebra of the partial observables $(x_1,p_1), (x_2,p_2)$. It is the relation between these and the Hamiltonian, or Hamiltonian constraint, that has the physical information about the system: not the Hamiltonian alone, or the Hamiltonian constraint alone. 

In the first case --two coupled harmonic oscillators whose evolution in an external time $t$ governed by the Hamiltonian\eqref{Hamop0}-- the Hamiltonian alone is incapable of distinguishing between a trivial system of two un-coupled oscillators, and the physically richer system of two coupled oscillators, with its specific phenomenology.

More radically, in the second case --two harmonic oscillators variables describing a single degree of freedom where the relative evolution of the two variables with respect to one another is governed by the Hamiltonian constraint \eqref{Ham0diagop}-- the Hamiltonian constrain alone does not provide any information at all about the physics (apart from the dimension of $\cal H$)!  All physical relevant information is coded into the transition amplitudes for the algebra of (partial) observables in $\cal K$.  

(Alternatively, and equivalently, the dynamics is in the evolving constants \cite{RovelliQM3}, which are Dirac observables well defined in $\cal H$:  these as well are not defined by the Hamiltonian constraint: they are determined by the partial observables in $\cal K$.)

One way or the other, it seems clear that the substantial part of a quantum theory is not restricted to dynamics: it is given by a specific choice of observables and the interplay between these and the dynamics. 

A set of subalgebras $\{ \mathcal{A}_i \}$, together with the constraint $C$ defined on a Hilbert space $\cal K$ contains the full necessary information to describe a quantum system. In that case, the minimal description of a quantum system will consist on prescribing the triple
$$
(\{ \mathcal{A}_i \}, C, \cal K).
$$
The observables that need to be specified are partial observables.

\section{Free considerations}

What is exactly the actual meaning of a TPS?  Is it a \emph{real} property of a system?  Do systems have a proper natural structure in components?  Is there such thing as ``the"  TPS of the system? Or does it depend on how we interact with the system? Is it a choice of the experimenter, namely, just a way for us to understand systems, to organize degrees of freedom and study them? Is it fundamental? Is there a canonical factorization of the world into subsystems? How fine is it? 

We are not going to answer these questions. We mention, for the sake of the discussion, the operational perspective suggested in \cite{Zanardi:2004tps}: the observables that induce TPS's are understood as those tunable and controllable by interactions the experimenter has access to.  There is a nice example of this in \cite{Zanardi:2004tps}: consider a spin singlet state in a 2-q-bit system.  If the experimental capabilities enable us to look at the Bell basis partition we can see entanglement between two components. But if we dispose only of a set of ``interactions and measurements'' that gives us access to the computational basis, the Bell states (maximal entangled states on the other basis) are product states in this basis, and no entanglement is present. One can therefore shift between TPS's by switching on and off different available interactions and measurements.

This perspective does not need to be anthropocentric as it sounds at first, since we can readily replace the ``experimenter" with an arbitrary second physical system, and consider the TPS as {\em relative} to this second system and its physical interaction Hamiltonian with the first. 

This is an interesting twist of the discussion, because it now seems again that it is the Hamiltonian that determines everything.  But the twist may be illusory, because the notion of interaction Hamiltonian requires the split between the ``observed" and the ``observing" system to be given first.  In other words, it requires a TPS to be given!  Certainly all physical relations are grounded on the dynamics, but the dynamics itself may be the dynamics of something. A simple vector in a Hilbert space may not be capable of supporting this `something'. 

The catch is general: quantum theory is a theory that describes how a system manifests itself to other systems, namely is a theory where a decomposition of the universe is assumed upfront.  This is explicit in many interpretations.  Certainly in the textbook Copenhagen interpretation, which splits the world into quantum system and classical apparatus.  Certainly in the Relational interpretation, where variables of one component take value only relative to  other components. And certainly in Q-bism, where the betting subject must distinguish itself from the system. But also in all Everettian interpretations of the Many Worlds kind, because the standard Everettian account of how variables come about {\em assumes} a split into subsystems.  The question of whether the decomposition of the universe in subsystems is primary, perspectival, determined by the dynamics, or else, seems unavoidable as soon as we remember that everything is likely to be described as quantum system, hence every system/observer split is a split of a quantum system. 

When we talk about a quantum {\em system} we are already assuming a split of the world in subsystems.

\section{Conclusions}

We have pointed out a number of difficulties in the idea of deriving the decomposition of a system in subsystems uniquely from the spectrum of the Hamiltonian.   

For normal systems with time, we have pointed out that we use Hamiltonians with the same spectrum to describe systems that we consider physically inequivalent: two coupled oscillators, or two uncoupled oscillators.  

More radically, we have argued that the physics of general covariant systems is encoded in the interplay between the dynamics defined by a Hamiltonian constraint and the partial observables.   This seems to suggest that the proper understanding of a quantum system requires the actual specification of the observable algebra.  

We have also pointed out that the decomposition of a system into subsystems represented by a tensorial decomposition of a Hilbert space is not only a fundamental global structure in many given quantum theories, but it plays also a more general structural role: the distinction between a special time variable is a TPS.  And the observer/observed split that is at the core of most interpretations of quantum theory are TPS as well.   


\bigskip\bigskip\bigskip
\noindent
{\bf{Acknowledgements}}~~
We thank I\~nigo L. Egusquiza for fruitful discussions when we started developing this work. We thank the Department of Theoretical Physics at UPV/EHU where most of this research was carried.
This
work was supported by the Basque Government Grant IT1628-22, and by the Grant PID2021-123226NB-I00
(funded by MCIN/AEI/10.13039/501100011033 and by ``ERDF A way of making Europe'').
We thank the whole Philosophy of Physics Reading Group of the Rotman Institute of Philosophy, and in particular Emily Adlam, Wayne Myrvold and Chris Smeenk, for their insights on a draft of this work. Finally, a deep thank to Carlo Rovelli for extensive comments and encouragements through the development of this work. 
FV's research at Western University is supported by the Canada Research Chairs Program, the QISS JFT grant 62312, and the Natural Science and Engineering Council of Canada (NSERC) through the Discovery Grant ``Loop Quantum Gravity: from Computation to Phenomenology". FV is affiliated to the Perimeter Institute for Theoretical Physics. Research at Perimeter Institute is supported by the Government of Canada through
Industry Canada and by the Province of Ontario through the Ministry of Economic Development and Innovation.
Western University and Perimeter Institute are located in the traditional lands of Anishinaabek, Haudenosaunee, L\=unaap\`eewak, Attawandaron and Neutral peoples. %
\bigskip




\end{document}